\RequirePackage{ifpdf}
\ifpdf 
\documentclass[pdftex]{sigma}
\else
\documentclass{sigma}
\fi

\begin{document}

\allowdisplaybreaks

\renewcommand{\PaperNumber}{034}

\renewcommand{\thefootnote}{$\star$}

\FirstPageHeading

\ShortArticleName{By Magri's Theorem, Self-Dual Gravity is Completely
Integrable}

\ArticleName{By Magri's Theorem, Self-Dual Gravity\\ is Completely
Integrable\footnote{This paper is a contribution to the Proceedings
of the Workshop on Geometric Aspects of Integ\-rable Systems
 (July 17--19, 2006, University of Coimbra, Portugal).
The full collection is available at
\href{http://www.emis.de/journals/SIGMA/Coimbra2006.html}{http://www.emis.de/journals/SIGMA/Coimbra2006.html}}}

\Author{Yavuz NUTKU}

\AuthorNameForHeading{Y. Nutku}

\Address{Feza G\"ursey Institute, P.O.Box 6, \c{C}engelk\"oy, Istanbul, 81220 Turkey}

\Email{\href{mailto:nutku@gursey.gov.tr}{nutku@gursey.gov.tr}, \href{mailto:notcoo@yahoo.com}{notcoo@yahoo.com}}

\ArticleDates{Received September 08, 2006, in f\/inal form
February 08, 2007; Published online February 27, 2007}

\Abstract{By Magri's theorem the bi-Hamiltonian structure of
Plebanski's second hea\-ven\-ly equation proves that (anti)-self-dual
gravity is a completely integrable system in four dimensions.}


\Classification{83C20; 83C15; 37K10}

\section{Introduction}
The purpose of this work is to present a recent result published in
my paper with Neyzi and Sheftel \cite{nns}, namely, the
bi-Hamiltonian structure \cite{magri1,magri2} of Plebanski's second
heavenly equation~\cite{pleb}.

Perhaps I may be allowed to start with a personal remark by way of
introduction. The f\/irst and the last time, before here in Coimbra,
that I had the pleasure of talking to Professor Magri was close to
20 years ago. I was just coming out of mid-life crisis
precipitated by having proved a very beautiful result in general
relativity \cite{nh} that I thought I could never equal again and
also divorce with very young children. I felt I had to change
everything. So I looked around and found shocks. Certainly what I
thought was the best developed subject in physics was the
symplectic structure underlying Newtonian mechanics. The great
works of Poisson, Hamilton, Jacobi, Liouville and Darboux two
centuries after Newton had made it a perfect gem. In particular
the theorem of Darboux stating that the symplectic $2$-form can be
put into the form given in the poster for this conference was the
f\/inal nail on the cof\/f\/in. Mind you this is not a local theorem
but, barring singularities, holds in all of phase space. All this
had made the theory of symplectic structure a fossilized subject
f\/it for Bourbaki!

Yet along came Nambu \cite{nambu} and Magri \cite{magri1,magri2} who
changed all that. In particular it was Magri who proved a very
beautiful and powerful theorem that an evolutionary system may admit
more than one Hamiltonian structure and was therefore completely
integrable. I found this most fascinating and started working on gas
dynamics which turned out to admit tri-Hamiltonian structure
\cite{gas} and later on was shown to admit one more \cite{gas2}.
However, nowadays most of the people in integrable systems refer to
\cite{gas2} instead of \cite{gas} where the original results f\/irst
appeared. I f\/ind it ironic that the people who originally shrugged
of\/f my results in \cite{gas} as being a trivial $2$-component system
in $1+1$ dimensions now use it as the cornerstone of their work
without supplying the correct reference.

So I was full of enthusiasm meeting Magri 20 years ago and wanted
to talk to him about his theorem. However, he just shrugged of\/f
and said that it was an old result. Now the theorem of Pythagoras
came two and a half millennia ago, have you ever seen any
reference to it as being old! Great theorems are never old and
Magri's theorem is a great one. On the other hand its author is
privileged to call it anything he likes.

Now I shall present you what I believe is the most important
application of Magri's theorem.

\section{Self-dual Gravity}

There are scalar-valued equations governing Riemannian metrics
with (anti)-self-dual Riemann $2$-form. By Bianchi's f\/irst
identity this implies Ricci-f\/latness. First we have the complex
Monge--Amp\`ere equation for K\"ahler metrics \cite{calabi}. This
has the form of the Monge--Amp\`ere determinant set equal to a
constant. Next we have the second heavenly equation of Plebanski
\cite{pleb}. They are related by a Legendre transformation which
in the context of Riemannian geometry is a~coordinate
transformation together with a redef\/inition of the K\"ahler
potential. The second heavenly equation of Plebanski, hereafter
denoted as $P_2$,
\begin{gather}
u_{tt} u_{xx} - u_{tx}^{2} + u_{xz} + u_{ty} = 0 ,
\label{heaven2}
\end{gather}
is close to the real Monge--Amp\`ere equation that has rich
Hamiltonian structure \cite{nrma2} and will be the subject of our
investigation.

\section[First order form of $P_2$]{First order form of $\boldsymbol{P_2}$}

The second heavenly equation is a second order partial
dif\/ferential equation. There is confusion in the literature as to
whether or not the independent variables in $P_2$ are complex, as
Plebanski originally presented, or real. We shall take them to be
real which leads to Euclidean signature for the metric. In order
to discuss its Hamiltonian structure we shall single out an
independent variable, $t$, in (\ref{heaven2}) to play the role of
``time" and express $P_2$ as a pair of f\/irst order nonlinear
evolution equations. Thus we introduce an auxiliary variable $q$
whereby (\ref{heaven2}) assumes the form
\begin{gather}
u_{t}  =    q ,   \qquad q_{t}  =
\frac{1}{u_{xx}} \left(
        q_{x}^{2}  - q_{y} - u_{xz} \right) \equiv Q
\label{uq}
\end{gather}
of a f\/irst order system. For the sake of brevity we shall
henceforth refer to (\ref{uq}) as the $P_2$-system. It is worth
noting that this split of (\ref{heaven2}) into the system
(\ref{uq}) is not unique, here we are using the most
straight-forward choice. Now the vector f\/ield
\begin{gather}
{\bf X} = q   \frac{\partial}{\partial u} + Q
 \frac{\partial}{\partial q}
\label{uqvector}
\end{gather}
def\/ines the f\/low. In the discussion of the Hamiltonian structure
of this system we shall use matrix notation with $ u^i$
$(i=1,2)$, $u^1=u$, $u^2 = q $ running over the dependent
variables.

The equations of motion (\ref{uq}) are to be cast into the form of
Hamilton's equations in two dif\/ferent ways according to the
recursion relation of Magri
\begin{gather*}
u^i_t = {\bf X}( u^i) = J^{ik}_{0}  \delta_k  H_{1} = J_1^{ik}
 \delta_k  H_{0}, 
\end{gather*}
where $\delta_k$ denotes variational derivative of the Hamiltonian
functional $H_i = \int_\infty^\infty {\cal H}_i dx dy dz$ with
respect to $u^k$.

\section{Lagrangian and Dirac's theory of constraints}

In order to arrive at the f\/irst Hamiltonian structure of the
$P_2$-system, we start with its Lagrangian
\begin{gather*}
{\cal L} = q  u_t u_{xx} + \frac{1}{2} u_t  u_y -
\frac{1}{2} q^2  u_{xx} + \frac{1}{2}  u_{x}  u_{z}
\end{gather*}
which is degenerate because its Hessian vanishes. Thus we need to
apply Dirac's theory of constraints \cite{dirac} in order to cast
it into Hamiltonian form. We def\/ine canonical momenta which
satisfy canonical Poisson brackets in the usual way. But we f\/ind
that they cannot be inverted for the velocities which must
therefore be imposed as constraints
\begin{gather*}
\phi_u  =  \pi_u - \left(q  u_{xx} + \frac{1}{2} u_y\right),\qquad
\phi_q =  \pi_q
\end{gather*}
and calculate the Poisson bracket of the constraints
\begin{gather}
K_{ik} = \left[ \phi_i(x,y,z) , \phi_k(x',y',z') \right]
= K_{ik}(x,y,z) \delta(x-x') \delta(y-y') \delta(z-z')
\label{kik}
\end{gather}
which plays an important role. The symplectic 2-form is obtained
by integrating the density
\begin{gather*}
\omega = \frac{1}{2}  d u^i \wedge K_{ij} d u^j
\end{gather*}
and it is straightforward to verify that the density of symplectic
$2$-form is determined by
\begin{gather*}
 K =    \left(            \begin{array}{ccc}
 q_{x} D_{x} + D_{x} q_{x} - D_{y}  & - u_{xx} \\
  u_{xx} & 0
\end{array}   \right)  
\end{gather*}
which is a f\/irst order local operator. Thus we f\/ind the symplectic
$2$-form
\begin{gather}
\omega =q_x  d u \wedge d u_{x} - u_{xx}
 d u \wedge d q - \frac{1}{2} d u \wedge d u_{y}
\label{omu}
\end{gather}
which, up to a divergence, can be directly verif\/ied to be a closed
2-form.

The statement of the symplectic structure of the equations of
motion (\ref{uq}) consists of
\begin{gather*}
 i_X \omega  =  d H   
\end{gather*}
which is obtained by the contraction of the closed symplectic
2-form $\omega$ with the vector f\/ield~{\bf X}~(\ref{uqvector})
def\/ining the f\/low.

\section{First Hamiltonian structure}

    The f\/irst Hamiltonian operator for $P_2$-system is
obtained by inverting (\ref{kik}) to arrive at the Dirac bracket.
This is given by
\begin{gather}
J_{0} = \left(            \begin{array}{cc}
 0 & \displaystyle\frac{ {   1} }{ {  u_{xx} } } \vspace{2mm}\\
\displaystyle- \frac{ {  1}}{  {  u_{xx} } }
 & \displaystyle   \frac{ {  q_x }}{{  u_{xx}^{2} }} D_x
+ D_x  \frac{ {  q_x }}{{  u_{xx}^{2} }} -
\frac{ {   1} }{ {  u_{xx} } } D_{y}
 \frac{ {   1} }{ {  u_{xx} } }
\end{array}  \right)               \label{j0}
\end{gather}
which, apart from the $D_y$ term in its last entry, is simply the
f\/irst Hamiltonian operator for the real Monge--Amp\`ere equation~\cite{nrma2}. It can be directly verif\/ied that
\begin{gather*}
{\cal H}_{1} = \frac{1}{2} q^2 u_{xx} -  \frac{1}{2} u_{x}
u_{z} 
\end{gather*}
is conserved for the f\/low~(\ref{uqvector}) and this is the f\/irst
Hamiltonian density appropriate to the operator~(\ref{j0}).

The proof  of the Jacobi identities for the Hamiltonian operator
(\ref{j0}) is straight-forward but rather lengthy. A shorter proof
follows from the observation that (\ref{j0}) is the inverse of
(\ref{kik}) which leads to the closed $2$-form (\ref{omu}).

\section{Recursion operator}
\label{sec-recurs}

We shall use the recursion operator in order to arrive at the
second Hamiltonian structure of the $P_2$-system. However, it is
worth recalling that the recursion operator originates in symmetry
analysis. So we start with the equation determining the symmetries
of the $P_2$-system and introduce two components for symmetry
characteristics
\begin{gather}
u_\tau = \varphi, \qquad q_\tau = \psi, \qquad
 \Phi \equiv \left(
\begin{array}{c} \varphi \\ \psi \end{array} \right)
\label{fipsi}
\end{gather}
while from the Frech\'et derivative of the f\/low we f\/ind
\begin{gather}
 {\cal A} =  \left( \begin{array}{cc} D_t & - 1 \vspace{2mm}\\ \displaystyle  \frac{{  Q }}
 {{  u_{xx} }} D_x^2 +
\frac{{  1 }}{{  u_{xx} }} D_x D_z
 & \displaystyle D_t - \frac{{  2 q_x }}{{  u_{xx} }} D_x
 + \frac{{  1 }}{{  u_{xx} }} D_y
 \end{array}
\right) \label{determiningop}
\end{gather}
so that the equation determining the symmetries of second heavenly
system is given by
\begin{gather}
 {\cal A}  \left( \Phi \right) = 0. \label{determiningeq}
\end{gather}
This determining equation has the divergence form
\begin{gather*}
(u_{xx}\psi-q_x\varphi_x+\varphi_y)_t +
(q_t\varphi_x-q_x\psi+\varphi_z)_x = 0 
\end{gather*}
which implies the local existence of the potential variable
$\tilde\varphi$ such that
\begin{gather}
\tilde\varphi_t = q_t\varphi_x-q_x\psi+\varphi_z ,\qquad
 \tilde\varphi_x = -(u_{xx}\psi-q_x\varphi_x+\varphi_y)
\label{pot}
\end{gather}
that satisf\/ies the same determining equation for symmetries of
(\ref{heaven2}) and therefore is a ``partner symmetry''
\cite{mnsbig} for (\ref{fipsi}). In the two-component form we
def\/ine the second component of this new symmetry similar to the
def\/inition of $\psi$ as $\tilde\psi = \tilde\varphi_t$. Then the
two-component vector
\[\tilde\Phi = \left(
\begin{array}{c} \tilde\varphi \\ \tilde\psi \end{array} \right)
\]
satisf\/ies the determining equation for symmetries in the form
(\ref{determiningeq}) and hence is a symmetry cha\-racteristic of
the system~(\ref{uq}) provided the vector~(\ref{fipsi}) is also a
symmetry characteristic. Thus~(\ref{pot}) become the recursion
relation for symmetries in the two-component form
\begin{gather*}
\tilde\Phi
 = {\cal R} ( \Phi )
\end{gather*}
with the recursion operator given by
\begin{gather}
 {\cal R} =  \left( \begin{array}{cc} D_x^{-1} (q_x D_x - D_y)
   & -  D_x^{-1} u_{xx}  \vspace{1mm}\\ Q D_x +  D_z
 &  - q_{x}
 \end{array}
\right),  \label{recursion}
\end{gather}
where $ D_{x}^{-1} $ is the inverse of $ D_{x}$. See
\cite{fokas} for the def\/inition and properties of this operator,
in particular,
\[ D_x^{-1} f =  \frac{1}{2}\left( \int^x_{-\infty} -
 \int_x^\infty\right)   f(\xi) \,  d \xi
 \] and the integrals
are taken in the principal value sense.

The commutator of the recursion operator (\ref{recursion}) and the
operator determining symmetries~(\ref{determiningop}) has the form
\begin{gather*}
\left[ {\cal R} ,  {\cal A} \right] = \left(
\begin{array}{cc}
D_x^{-1}(q_t-Q)_{xx}-(q_t-Q)_x & D_x^{-1}(u_t-q)_{xx}
\vspace{1mm}\\
\displaystyle \left\{\frac{  Q}{
u_{xx}}(u_t-Q)_{xx}+\left(D_y-\frac{  2q_x}{  u_{xx}}
\right)({  q_t-Q })_x\right\}D_x & (q_t-Q)_x
\end{array}
\right) 
\end{gather*}
and as a consequence, the operators ${\cal R}$ and ${\cal A}$
commute
\begin{gather*}
 \left[ {\cal R} ,  {\cal A} \right] = 0
\end{gather*}
by virtue of the $P_2$-system (\ref{uq}). Moreover, $ {\cal R}$
and $ {\cal A} $ form a Lax pair for the second heavenly system.

\section{Second Hamiltonian structure}
\label{sec-multihamilton}

   The second Hamiltonian operator $J_1$ is obtained by applying the
recursion operator (\ref{recursion}) to the f\/irst Hamiltonian
operator $J_1 = {\cal R} J_0$. We f\/ind
\begin{gather}
J_{1} =   \left(            \begin{array}{cc}  D_{x}^{-1}
 & \displaystyle - \frac{{  q_x }}{{  u_{xx} }} \vspace{2mm}\\
 \displaystyle \frac{{  q_x }}{{  u_{xx} }} &
 \begin{array}{c}
 \displaystyle -\frac{1}{2} \left(Q D_x  \frac{{  1
}}{{  u_{xx} }} + \frac{{  1 }}{{
u_{xx} }} D_x Q \right)
\vspace{1mm}\\ \displaystyle{}+ \frac{1}{2} \left( \frac{{  q_x }}{{  u_{xx} }}
D_y  \frac{{  1 }}{{  u_{xx} }} +
\frac{{  1 }}{{  u_{xx} }} D_y \frac{{
q_x }}{{  u_{xx} }}
 + \frac{{  1 }}{{  u_{xx} }}
D_z + D_z \frac{{  1 }}{{  u_{xx} }}  \right)
\end{array}
\end{array}  \right)               \label{j1}
\end{gather}
which is manifestly skew. The proof of the Jacobi identity is
again straight-forward and lengthy.

The Hamiltonian operators (\ref{j0}) and (\ref{j1}) form a Poisson
pencil, that is, every linear combination $aJ_0+bJ_1$ of these two
Hamiltonian operators with constant coef\/f\/icients $a$ and $b$
satisf\/ies the Jacobi identities.

\section{Conclusion}

The second heavenly equation of Plebanski which determines
Riemannian metrics with (anti)-self-dual curvature $2$-form
implying Ricci-f\/latness admits bi-Hamiltonian structure. Therefore
by Magri's theorem self-dual gravity is a completely integrable
system in 4 dimensions.

\pdfbookmark[1]{References}{ref}
\LastPageEnding

\end{document}